\documentstyle[twocolumn,aps]{revtex}
\newcommand{\half}{\frac{1}{2}}
\begin{document}
\draft
\title{
Reply to Lindgren and Salomonson}
\author{R. K. Nesbet}
\address{
IBM Almaden Research Center,
650 Harry Road,
San Jose, CA 95120-6099, USA}
\date{\today}
\maketitle
\begin{center}For {\em Phys.Rev.A} \end{center}
\begin{abstract}
In the accompanying Comment [Phys. Rev. A {\bf 67}, 056501 (2003)],
I. Lindgren and S. Salomonson claim to
prove for the Kohn-Sham kinetic energy functional of ground state 
electron density that a Fr\"echet functional derivative exists, 
equivalent to a multiplicative local potential function.  If true, this 
result would imply an exact Thomas-Fermi theory for ground states of 
noninteracting electrons.  However, such a theory is not consistent with
the exclusion principle for more than one electron of each spin.  The
simplest counterexample is the lowest triplet state of a noninteracting 
two-electron atom.  If only the total electron density were normalized, 
as in Thomas-Fermi theory, the lowest state would collapse into a 
doubly-occupied $1s$ spin-orbital.  Two independent parameters 
$\epsilon_{1s}$ and $\epsilon_{2s}$ are required to maintain independent
subshell normalization.  The argument presented by these authors is 
discussed in the light of this unphysical implication.
\end{abstract}
\vspace*{5mm}
\par
The underlying issue here is whether Hohenberg-Kohn theory\cite{HAK64}
implies an exact {\it ab initio} Thomas-Fermi theory (TFT) for ground 
states.  For noninteracting electrons, the question reduces to the 
existence of a Fr\'echet functional derivative\cite{BAB92} of the 
Kohn-Sham kinetic energy functional\cite{KAS65}, equivalent to a 
multiplicative local potential function.  If such a functional 
derivative did exist, it could be expressed as
\begin{eqnarray}\label{TFTeq}
\frac{\delta T}{\delta\rho}=\mu-v({\bf r}),
\end{eqnarray}
which simply states the TFT Euler-Lagrange equation.  Constant $\mu$ is 
determined by the normalization condition $\int\rho=N$ for N electrons. 
In the context of density functional theory (DFT),
$\rho=\sum_in_i\rho_i=\sum_in_i\phi^*_i\phi_i$,
a sum of orbital subshell densities.  Density variations generated by 
unconstrained independent variations in the orbital Hilbert space are 
necessary and sufficient to determine the Euler-Lagrange equations.
Spin indices and sums are assumed but suppressed in the notation here. 
For noninteracting electrons, the correct Euler-Lagrange equations for 
Fermi-Dirac electrons, as used in Kohn-Sham theory, are a set of 
Schr\"odinger equations, 
\begin{eqnarray}\label{DFTeq}
{\hat t}\phi_i({\bf r})=\{\epsilon_i-v({\bf r})\}\phi_i({\bf r}),
\end{eqnarray}
where ${\hat t}=-\half\nabla^2$, for orthogonal but independently 
normalized orbital functions $\phi_i({\bf r})$.  Independent orbital 
normalization requires independent parameters $\epsilon_i$, and orbital
orthogonality implies the exclusion principle.  The restriction to one
free parameter in Eq.(\ref{TFTeq}) means that the exclusion principle 
cannot be enforced unless all energy eigenvalues $\epsilon_i$ are equal.
Hence, in a compact system, these equations cannot be reconciled for 
more than one electron of each spin.  The simplest specific example is
the lowest $1s2s\;^3S$ state of a two-electron system.  TFT implies 
bosonic condensation with two electrons in the $1s$ spin-orbital.  The 
argument of Lindgren and Salomonson\cite{LAS03} (LS) must fail, because
it implies that TFT is an exact theory for noninteracting electrons. 
The proposed argument also disagrees with DFT, in which 
the Kohn-Sham equations for noninteracting electrons are just 
Eqs.(\ref{DFTeq}), valid for atomic shell structure and consistent with 
the exclusion principle.  The TFT and DFT Euler-Lagrange equations are 
inconsistent in the $1s2s\;^3S$ example, because two independent
constants cannot be deduced from one. This conclusion has been disputed 
and defended in recent publications\cite{GAL00,HAM01,NES02}.  
\par 
LS put forward two variational derivations that set out to 
determine the density functional derivative of the Kohn-Sham
kinetic energy functional $T_s[\rho]$, constructed as the
ground-state limit of the Schr\"odinger orbital functional
$T[\{\phi_i\}]=\sum_in_i(i|{\hat t}|i)$.  In Section IIA,
they consider an N-electron wave function $\Psi$.  For noninteracting
electrons, the Schr\"odinger equation is separable, and the wave 
function can be expressed as a serial product of orbital functions,
$\Psi=\Pi_i \phi_i({\bf r_i})$.  Fermi-Dirac statistics are not implied,
but must be imposed by explicit orthogonalization and by independent
normalization of the partial densities $\rho_i=\phi^*_i\phi_i$.
In terms of partial densities, LS Eq.(18) should be
$\delta T=\int\sum_in_i(\epsilon_i-v)\delta\rho_i$. 
Instead, LS extend the functional derivative of Eq.(22) 
to unnormalized densities by making the 
implicit assumption that the undetermined constant has no orbital
subshell structure, which is inconsistent with the orbital
Schr\"odinger equations.  This point can be examined in detail by 
considering specific wave functions, as in LS Section IIB.
\par 
In Section IIB, LS propose a variant of standard Hartree-Fock theory in
which the product of orbital overlap integrals is retained, in contrast 
to the usual simplified argument based on the orbital functional $T$.
If done correctly, this variant
leads to standard Hartree-Fock equations, equivalent for noninteracting
electrons to Kohn-Sham equations, in which the kinetic energy operator
${\hat t}$ takes its usual form.  The problem here is that all terms
in the total energy expression for a Slater determinant must be 
considered together.  To enforce orbital orthonormality, as required by 
the definition of a Slater determinant, a matrix of Lagrange multipliers
$\epsilon_{ij}$ is required.  It is customary (for closed-shell systems)
to assume a diagonalized canonical representation, leaving the diagonal 
elements $\epsilon_i$ to be determined as eigenvalues
of the canonical Hartree-Fock 
equations.  These are Lagrange multipliers for independent normalization
constraints $(i|i)=1$.  Derivations using the simplified orbital 
functional $T$ are standard in applications of DFT and are consistent 
with Hartree-Fock theory.  Products of the overlap factors are unity for
variational solutions and can consistently be omitted. If retained in 
the derivation, overlap variations are multiplied by a factor that 
vanishes.  An appropriate variational functional for noninteracting 
electrons is 
\begin{eqnarray}\label{fprod}
\sum_in_i(i|{\hat t}+v({\bf r})-\epsilon_i|i)\Pi_{j\neq i}(j|j).
\end{eqnarray}
This is derived directly for a serial product wave function, with
Lagrange multipliers for independent orbital normalization.  Orbital
orthogonality is implied by the Euler-Lagrange equations or by symmetry.
The first factor in each term here vanishes in stationary states,
so all variations of the overlap product drop out, and variations of 
the first factor are multiplied by unity.  Written out in full, the 
correct differential for variations about a stationary state is
\begin{eqnarray}\label{dprod}
\sum_in_i&&[\int d^3{\bf r}
\delta\phi^*_i({\bf r})\{{\hat t}+v({\bf r})-\epsilon_i\}\phi_i({\bf r})
\Pi_{j\neq i}(j|j)+cc ]
\nonumber\\
+\sum_in_i&&
(i|{\hat t}+v({\bf r})-\epsilon_i|i)\delta\Pi_{j\neq i}(j|j). 
\end{eqnarray}
This vanishes for a stationary state, but cannot do so unless all 
Lagrange multipliers $\epsilon_i$ are included, precluding an equivalent
Thomas-Fermi theory. 
\par The argument presented by LS fails to consider the full 
variational expression including $(i|v({\bf r})-\epsilon_i|i)$ for each 
occupied orbital.  LS Eq.(33) should be written as
\begin{eqnarray}
\delta T_s&=&\langle\delta\phi_1|\epsilon_1-v({\bf r})|\phi_1\rangle
\langle 2|2 \rangle
\nonumber\\
&+&\langle\delta\phi_2|\epsilon_2-v({\bf r})|\phi_2\rangle
\langle 1|1 \rangle+cc ,
\end{eqnarray}
plus terms involving $\delta(i|i)$ that drop out of Eqs.(\ref{dprod}).
A correct derivation must take into account that the normalization
product modifies all terms of Eq.(\ref{fprod}), and modifies the usual 
DFT definition $\rho=\sum_in_i\rho_i$.  The functional derivative must 
be consistent with the Hartree-Fock equations implied for Fermi-Dirac
electrons.  Failure to include a separate Lagrange multiplier for
normalization of each occupied orbital is inconsistent with the 
constraints needed to specify a Slater determinant as the variational 
trial function.  
\par
The chain rule connecting orbital and density functional derivatives of
the functional $T_s[\rho]$ is a consistency condition implied by the 
Kohn-Sham construction of $T_s$ using the Schr\"odinger orbital
functional $T$\cite{NES02}.
In their discussion of the chain rule appropriate to DFT, LS invoke
an expression, LS Eq.(36), that is undefined unless a Fr\'echet
functional derivative exists.  The integral in LS Eq.(36) is
inappropriate, since $\rho$ is an explicit function of $\phi^*_i$.  
LS Eq.(38) is inconsistent with the orbital Euler-Lagrange equations.
Such logic can be avoided because the ground-state functional,
extended to include density variations unrestricted by normalization,
defines G\^ateaux derivatives\cite{BAB92} such that\cite{NES02} 
\begin{eqnarray}
\frac{\delta T_s}{n_i\delta\rho_i}\phi_i=
\frac{\delta T_s}{n_i\delta\phi^*_i}={\hat t}\phi_i=
\{\epsilon_i-v\}\phi_i.
\end{eqnarray}
A Fr\'echet derivative is implied only if all $\epsilon_i$ are equal.
Alternatively, since these G\^ateaux derivatives are operationally
equivalent to the Schr\"odinger operator ${\hat t}$, one can define
a generalized functional derivative as such a linear operator.  Using 
this generalized definition, the Kohn-Sham equations can be derived 
entirely in terms of density functional derivatives\cite{NES98}.
\par
LS misstate the mathematical implications of the fact that the
noninteracting Kohn-Sham or Schr\"odinger equations imply by direct
construction the existence of G\^ateaux functional derivatives of the
extended functional $T_s$\cite{NES02}.  Unless all $\epsilon_i$ are 
equal, this implies that a Frech\'et functional derivative does not 
exist.  If a Fr\'echet derivative is simply assumed, in order to define 
an extended functional of total density, the theory is inconsistent with
Fermi-Dirac statistics.  As a consequence of the exclusion principle, 
there is no physically correct definition of such an extended 
functional.  However, because G\^ateaux derivatives exist, 
a unique extended functional of the subshell densities $\rho_i$ is
defined.  This suffices to determine the Euler-Lagrange equations of
the theory\cite{NES02a}.  For practical applications
of DFT, which must assume a parametrized functional $E_{xc}[\rho]$, the
implied universal functional $F_s$ has the same properties as $T_s$.  An
extended functional $F_s[\rho]$ and the Fr\'echet derivative required by
TFT do not exist in general for more than two electrons, but a unique 
extended functional of the partial densities $\rho_i$ does exist, as do 
the G\^ateaux derivatives required to determine Euler-Lagrange 
equations.  This generalizes TFT to a theory consistent with Fermi-Dirac
statistics and with electronic shell structure\cite{NES02a}.

\end{document}